\documentclass[12pt]{article}
\begin{document}

\title{\bf The Structure of the Pseudo-Newtonian Force and Potential
about a Five Dimensional Rotating Black Hole}

\author{M. Sharif \thanks{e-mail: msharif@math.pu.edu.pk}
\\ Department of Mathematics, University of the Punjab,\\
Quaid-e-Azam Campus Lahore-54590, PAKISTAN.}

\date{}

\maketitle
\begin{abstract}
In this paper we investigate the structure of the pseudo-Newtonian
force and potential about a five dimensional rotating black hole.
The conditions for the force character from an attractive to
repulsive are considered. It is also found that the force will
reach a maximum under certain conditions.
\end{abstract}

{\bf Keywords: Force and Potential, Five Dimensional Rotating
Black Holes}


Einstein's Theory of Relativity replaces the use of forces in
dynamics by what Wheeler calls {\it geometrodynamics}. Paths are
bent, not by forces, but by the {\it curvature of spacetime}. In
the process the guidance of the intuition based on the earlier
dynamics is lost. However, our intuition continues to reside in
the force concept, particularly when we have to include other
forces in the discussion. For this reason one may want to reverse
the procedure of General Relativity and look at the non-linear
{\it force of gravity} which would predict the same bending of the
path as predicted by geometry. In the pseudo-Newtonian ($\psi N$)
approach [1,2], the curvature of the spacetime is {\it
straightened out} to yield a relativistic force which bends the
path, so as to again supply the guidance of the earlier,
force-based, intuition.

The relativistic analogue of the Newtonian gravitational force
which gives the relativistic expression for the tidal force in
terms of the curvature tensor is called the $\psi N$ force. The
quantity whose gradient gives the $\psi N$ force is the $\psi N$
potential. The $\psi N$ force is the generalisation of the force
which gives the usual Newtonian force for the Schwarzschild metric
and a $"\frac{Q^2}{r^3}"$ correction to it in the
Riessner-Nordstrom metric. The $\psi N$ force may be regarded as
the {\it Newtonian fiction} which {\it explains} the same motion
(geodesic) as the {\it Einsteinian reality} of the curved
spacetime does. We can, thus, translate back to Newtonian terms
and concepts where our intuition may be able to lead us to ask,
and answer, questions that may not have occurred to us in
relativistic terms. Some insights have already been obtained
[1]-[4] by expressing the consequences of General Relativity in
terms of forces by applying it to Kerr and Kerr-Newmann metrics.
Recently, the pseudo-Newtonian potential has been evaluated for
charged particle in Kerr-Newmann geometry by Ivanov and Prodanov
[5]. In this paper, we study the structure of force and potential
about a five dimensional rotating black hole.

The basis of the $\psi N$ formalism is the observation that,
whereas the gravitational force is not detectable in a freely
falling frame, {\it that is so only at a point}. It is detectable
over a finite spatial extent as the tidal force. The tidal force,
which is operationally determinable, can be related to the
curvature tensor by
\begin{equation}
F_T^\mu=mR_{\nu\rho\pi}^\mu t^\nu l^\rho t^\pi,\quad
(\mu,\nu,\rho,\pi=0,1,2,3),
\end{equation}
where $m$ is the mass of a test particle, $t^\mu=f(x)\delta_0^\mu,
\quad f(x)=(g_{00})^{-1/2}$ and $l^\mu$ is the separation vector.
$l^\mu$ can be determined by the requirement that the tidal force
have maximum magnitude in the direction of the separation vector.
Choosing a gauge in which $g_{0i}=0$ (similar to the synchronous
coordinate system [6,7]) in a coordinate basis. We further use
Riemann normal coordinates for the spatial direction, but not for
the temporal direction. The reason for this difference is that
both ends of the accelerometer are spatially free, i.e., both move
and do not stay attached to any spatial point. However, there is a
{\it memory} of the initial time built into the accelerometer in
that the zero position is fixed then. Any change is registered
that way. Thus {\it time} behaves very differently from {\it
space}.

The $\psi N$ force, $F_\mu$, satisfies the equation
\begin{equation}
F_T^{*\mu}=l^\nu F_{;\nu}^\mu ,
\end{equation}
where $F_T^{*\mu}$ is the extremal tidal force corresponding to
the maximum magnitude reading on the dial. Notice that
$F_T^{*0}=0$ does not imply that $F^0=0$. The requirement that
Eq.(2) be satisfied can be written as
\begin{equation}
l^i(F_{,i}^0+\Gamma_{ij}^0F^j)=0,
\end{equation}
\begin{equation}
l^j(F_{,j}^i+\Gamma_{0j}^iF^0)=F_T^{*i},\quad (i,j=1,2,3).
\end{equation}
A simultaneous solution of the above equations can be found by
taking the ansatz
\begin{equation}
F^0=m\left[(\ln
A)_{,0}-\Gamma_{00}^0+\Gamma_{0j}^i\Gamma_{0i}^i/A\right]f^2,
\end{equation}
\begin{equation}
F^i=m\Gamma_{00}^i f^2,
\end{equation}
where $A=(\ln \sqrt{-g})_{,0},\quad g=det(g_{ij})$. These
equations can be written in terms of two quantities $U$ and $V$
given by
\begin{equation}
U=m\left[\ln (Af/B)-\int(g^{ij}_{,0} g_{ij,0}/4A)dt\right],
\end{equation}
\begin{equation}
V=-m\ln f,
\end{equation}
as
\begin{equation}
F_0=-U_{,0}, \quad F_i=-V_{,i},
\end{equation}
where $B$ ia a constant with units of time inverse so as to make
$A/B$ dimensionless. It is to be noted that the momentum
four-vector $p_\mu$ can be written in terms of the integral of the
force four-vector $F_\mu$. Thus
\begin{equation}
p_{_\mu }=\int F_\mu d\tau.
\end{equation}
Notice that the zero component of the momentum four-vector
corresponds to the energy imparted to a test particle of mass $m$
while the spatial components give the momentum imparted to a test
particle.

Thus, in the free fall rest-frame, the $\psi N$ force is given
[2,3] by
\begin{equation}
F_i=-m(\ln \sqrt {g_{00}})_{,i},\quad =-V_{,i},
\end{equation}
where
\begin{equation}
V=m(\ln \sqrt {g_{00}}).
\end{equation}
It is clear that $V$ is the generalisation of the classical
gravitational potential and, for small variations from Minkowski
space
\begin{equation}
V\approx\frac{1}{2}m(g_{00}-1)
\end{equation}
which is the pseudo-Newtonian potential. We shall analyse the
behaviour of these quantities for the five dimensional rotating
black hole.

The metric of a rotating black hole in five dimensions follows
from the general asymptotically flat solutions to $(N+1)$
dimensional vacuum gravity found by Myers and Perry [8]. In
Boyer-Lindquist type coordinates, it takes the most simple form
given by
\begin{eqnarray}
ds^2&=&-dt^2+\Sigma(\frac{r^2}{\Pi}dr^2+d\theta^2)+(r^2+a^2)\sin^2
\theta d\phi^2+(r^2+b^2)\cos^2\theta d\psi^2\nonumber\\
&+&\frac{M}{\Sigma}(dt-a\sin^2\theta d\phi-b\cos^2\theta d\psi)^2,
\end{eqnarray}
where
\begin{equation}
\Sigma=r^2+a^2\cos^2\theta+b^2\sin^2\theta, \quad
\Pi=(r^2+a^2)(r^2+b^2)-Mr^2,
\end{equation}
and $M$ is a parameter related to the physical mass of the black
hole, while the parameters $a$ and $b$ are associated with its two
independent angular momenta.

The event horizon of the black hole is a null surface determined
by the equation
\begin{equation}
\Pi=(r^2+a^2)(r^2+b^2)-Mr^2=0.
\end{equation}
The largest root of this equation gives the radius of the black
hole's outer event horizon. We have
\begin{equation}
r^2_h=\frac{1}{2}(M-a^2-b^2+\sqrt{(M-a^2-b^2)^2-4a^2b^2}.
\end{equation}
Notice that the horizon exists if and only if
\begin{equation}
a^2+b^2+2|ab|\leq M,
\end{equation}
so that the condition $M=a^2+b^2+2|ab|$ or, equivalently,
$r^2_h=|ab|$ defines the extremal horizon of a five dimensional
black hole.

In the absence of the black hole $(M=0)$, the metric (14) reduces
to the flat one written in oblate bi-polar coordinates. On the
other hand, for $a=b=0$ we have the Schwarzschild-Tangherlini [9]
static solution in spherical bi-polar coordinates. The
bi-azimuthal symmetry properties of the five dimensional black
hole metric (14) along with its stationarity imply the existence
of the three commuting Killing vectors
\begin{equation}
\xi_{(t)}=\frac{\partial}{\partial t}, \quad
\xi_{(\phi)}=\frac{\partial}{\partial \phi}, \quad
\xi_{(\psi)}=\frac{\partial}{\partial \psi}.
\end{equation}
The scalar products of these Killing vectors are expressed through
the corresponding metric components as follows
\begin{eqnarray}
\xi_{(t)}.\xi_{(t)}&=&g_{tt}=-1+\frac{M}{\Sigma},\nonumber\\
\xi_{(\phi)}.\xi_{(\phi)}&=&g_{\phi\phi}
=(r^2+a^2+\frac{Ma^2}{\Sigma}\sin^2\theta)\sin^2\theta,\nonumber\\
\xi_{(t)}.\xi_{(\phi)}&=&g_{t\phi}=-\frac{Ma}{\Sigma}\sin^2\theta,\nonumber\\
\xi_{(\psi)}.\xi_{(\psi)}&=&g_{\psi\psi}
=(r^2+b^2+\frac{Mb^2}{\Sigma}\cos^2\theta)\cos^2\theta,\nonumber\\
\xi_{(t)}.\xi_{(\psi)}&=&g_{t\psi}=-\frac{Mb}{\Sigma},\nonumber\\
\xi_{(\phi)}.\xi_{(\psi)}&=&g_{\phi\psi}
=\frac{Mab}{\Sigma}\sin^2\theta\cos^2\theta.
\end{eqnarray}
The Killing vectors (19) can be used to give a physical
interpretation of the parameters $M,~a$ and $b$ involved in the
metric (14). One can obtain coordinate-independent definitions for
these parameters by using the analysis given in [10,11]. We have
the integrals
\begin{equation}
M=\frac{1}{4\pi^2}\oint{\xi^{\mu;\nu}_{(t)}}d^3\Sigma_{\mu\nu},
\end{equation}
\begin{equation}
j(a)=aM=-\frac{1}{4\pi^2}\oint{\xi^{\mu;\nu}_{(\phi)}}d^3\Sigma_{\mu\nu},\quad
j(b)=bM=-\frac{1}{4\pi^2}\oint{\xi^{\mu;\nu}_{(\psi)}}d^3\Sigma_{\mu\nu},
\end{equation}
where the integrals are taken over the 3-sphere at spatial
infinity,
\begin{equation}
d^3\Sigma_{\mu\nu}=\frac{1}{3!}\sqrt{-g}\varepsilon_{\mu\nu\alpha\beta\gamma}
dx^\alpha\wedge dx^\beta\wedge dx^\gamma,
\end{equation}
the semicolon denotes covariant differentiation. The two specific
angular momentum parameters $j(a)$ and $j(b)$ are associated with
rotations in the $\phi$ and $\psi$ directions respectively. It is
mentioned here that with these definitions the relation between
the specific angular momentum and the mass parameter looks exactly
like the corresponding relation ($J = aM_T$) of four dimensional
Kerr metric. It can be shown that the definitions given in (21)
and (22) do in fact correctly describe the mass and angular
momenta parameters. The integrands can be calculated in the
asymptotic region $r\rightarrow\infty$. The dominant terms in the
asymptotic expansion have the form
\begin{eqnarray}
\xi_{(t)}^{t;r}&=&\frac{M}{r^3}+O(\frac{1}{r^5}),\nonumber\\
\xi_{(\phi)}^{t;r}&=&-\frac{2aM\sin^2\theta}{r^3}+O(\frac{1}{r^5}),\nonumber\\
\xi_{(\psi)}^{t;r}&=&-\frac{2bM\cos^2\theta}{r^3}+O(\frac{1}{r^5}).
\end{eqnarray}
One can easily verify that these expressions satisfy the formulae
(21) and (22). On the other hand, the relation of the above
parameters to the total mass $M_T$ and the total angular momenta
$J(a)$ and $J(b)$ of the black hole can be established using the
formulae given in [8]. We obtain that
\begin{equation}
M=\frac{8}{3\pi}M_T,\quad j(a)=\frac{4}{\pi}J(a),\quad
j(b)=\frac{4}{\pi}J(b).
\end{equation}
These relations confirm the interpretation of the parameters $M,
a$ and $b$ as being related to the physical mass and angular
momenta of the metric (14).

The structure of the $\psi N$ force (per unit mass of the test
particle) for the five dimensional rotating black hole turns out
to be the following
\begin{equation}
F_r=-\frac{Mr}{\Sigma(\Sigma-M)},
\end{equation}
\begin{equation}
F_\theta=\frac{M(a^2-b^2)\sin\theta\cos\theta}{\Sigma(\Sigma-M)}.
\end{equation}
We see that the radial component can never become zero outside the
horizon and so the force cannot change character from an
attractive to a repulsive force outside the black hole. The polar
component can only become zero outside the horizon at
$\theta=0,~\pi/2,~\pi$. We note that naked singularities can give
repulsive as well as attractive forces. The force structure can
provide interesting features if it reaches a maximum and then
drops as we reduce $r$ or change $\theta$ provided the turnover
lies outside the horizon. Since our observers are seeing {\it
force} in a flat space, the metric to be used is the plane polar
one. Thus the square of the magnitude of the force is
\begin{equation}
(F)^2=\frac{M^2}{r^2\Sigma^2(\Sigma-M)^2}[r^4
+(a^2-b^2)^2\sin^2\theta\cos^2\theta].
\end{equation}
It follows from here that
\begin{equation}
|F|=\frac{M}{r\Sigma(\Sigma-M)}[r^4
+(a^2-b^2)^2\sin^2\theta\cos^2\theta]^{1/2}.
\end{equation}
The expansion of $|F|$ in powers of $(1/r)$ can be given as
\begin{equation}
|F|=\frac{Mr}{\Sigma(\Sigma-M)}[1+\frac{(a^2-b^2)^2
\sin^2\theta\cos^2\theta}{2r^4}+....].
\end{equation}
This shows that the force far from the center, i.e., when $r$ is
very very large, becomes zero. This corresponds to the behaviour
of the Kerr metric [4].

The equations for the turnovers along $r$ and $\theta$,
respectively, are
\begin{eqnarray}
r^4[\Sigma(\Sigma-M)-2r^2(2\Sigma-M)]
-(a^2-b^2)^2[\Sigma(\Sigma-M)\nonumber\\
+2r^2(2\Sigma-M)]\sin^2\theta\cos^2\theta=0,
\end{eqnarray}
\begin{equation}
\Sigma(\Sigma-M)(a^2-b^2)\cos2\theta+2(2\Sigma-M)
[r^4+(a^2-b^2)^2\sin^2\theta\cos^2\theta]=0.
\end{equation}
Eqs.(31) and (32) are not easy to analyse generally. We
investigate these equations for the following two special cases.\\
\par \noindent
(i)~$a=b\neq0$,\quad (ii)~$a=0=b$.\\
\par \noindent
For the first case (i), when the two angular momenta are the same,
Eqs.(31) and (32) reduce to
\begin{equation}
(r^2+a^2)(r^2+a^2-M)-2r^2(2r^2+2a^2-M)=0,
\end{equation}
\begin{equation}
2(r^2+a^2)-M=0.
\end{equation}
It follows that the first equation is satisfied when
\begin{equation}
r^2=\frac{1}{6}[(M-2a^2)\pm\sqrt{M^2+16a^4-16Ma^2}].
\end{equation}
Eq.(34) is satisfied for the value of $r$ given by
\begin{equation}
r^2=\frac{1}{2}M-a^2.
\end{equation}
If we take $a=0=b$, i.e., there is no rotation, the values of $r$
satisfying Eqs.(31) and (32), respectively, are
\begin{equation}
r=\pm\sqrt{\frac{M}{3}},\quad r=\pm\sqrt{\frac{M}{2}}.
\end{equation}
We note that in the first case, a maximum of the magnitude does
occur at the value of $r$ given by Eq.(35). The maximum value of
the force can be obtained by replacing this value of $r$ in $F_r$.
Similarly, for the case (ii), a maximum value of the magnitude is
obtained for $r$ as given in Eq.(37).

The corresponding potential is obtained by using Eqs.(13) and (14)
and is given by
\begin{equation}
V=-\frac{m}{2\Sigma}.
\end{equation}
For the special case (i), it becomes
\begin{equation}
V=-\frac{m}{2(r^2+a^2)}.
\end{equation}
When there is no rotation it reduces to
\begin{equation}
V=-\frac{m}{2r^2}.
\end{equation}
We note that the structure of force and potential indicate similar
type of behaviour as for the Kerr metric [4].



\vspace{2cm}

{\bf \large References}

\begin{description}

\item{[1]} Qadir A. and Quamar J.: {\it Proc. 3rd Marcel Grossmann Meeting on
General Relativity}, ed. Hu Ning (North Holland Amstderm 1983)189;\\
Quamar J.: {\it Ph.D. Thesis} Quaid-i-Azam University Islamabad
(1984).

\item{[2]} Qadir Asghar and Sharif M.: Nuovo Cimento B {\bf 107}(1992)1071;\\
Sharif M.: {\it Ph.D. Thesis} Quaid-i-Azam University Islamabad
(1991).

\item{[3]} Qadir Asghar and Sharif M.: Phys. Lett. A {\bf 167}(1992)331;\\
Sharif M.: Astrophys. and Space Science {\bf 253}(1997)195.

\item{[4]} Qadir A. and Quamar, J.: Europhys. Lett. {\bf
2}(1986)423;\\
Qadir A.: Europhys. Lett. {\bf 2}(1986)427.

\item{[5]} Ivanov, Rossen I. and Prodanov, M.: Phys. Lett. B {\bf611}(2005)34.

\item{[6]} Misner, C.W., Thorne, K.S. and Wheeler, J.A.: {\it Gravitation}
(W.H. Freeman San Francisco, 1973).

\item{[7]} Landau, L.D. and Lifschitz, E.M.: {\it The Classical Theory of
Fields} (Pergamon Press, 1975).

\item{[8]} Myers, R.C. and Perry, M.J.: Ann. Phys. {\bf172}(1986)304.

\item{[9]} Tangherlini, F.R.: Nuovo Cimeto {\bf B27}(1963)636.

\item{[10]} Aliev, A.N. and  Frolov, Valeri, P.: Phys. Rev. {\bf D69}(2004)084022.

\item{[11]} Komar, A.: Phys. Rev. {\bf 113}(1959)934.

\end{description}

\end{document}